\documentclass{IEEEtran}
\usepackage[font=footnotesize,labelfont=bf]{caption}
\IEEEoverridecommandlockouts

\usepackage{cite}
\usepackage{amsmath,amssymb,amsfonts}
\usepackage{algorithmic}
\usepackage{graphicx}
\usepackage{textcomp}
\usepackage{url}
\usepackage{xcolor}
\usepackage{array}
\usepackage{longtable}
\usepackage{booktabs}
\usepackage{tabularx}
\usepackage{tabularray}
\usepackage{caption} 

\usepackage[frozencache=true,cachedir=minted-cache]{minted} 
\usepackage[most]{tcolorbox}
\usepackage{xcolor} 
\usepackage{balance}
\usepackage{pgfplots}
\pgfplotsset{compat=1.17}
\usepackage{float}
\usepackage[bookmarks=false]{hyperref}

\usepackage{pdfpages}
\pdfoutput=1
\usepackage{hyperref}
\hypersetup{
  pdfinfo={
    Title={Decoding Dependency Risks: A Quantitative Study of Vulnerabilities in the Maven Ecosystem},
    Author={Costain Nachuma},
    Subject={MSR 2025},
    Keywords={dependency vulnerability, maven central, software ecosystem, open source software}
  }
}
\usepackage{amssymb}

\definecolor{lightblue}{RGB}{180, 219, 248} 

\tcbset{
    findingstyle/.style={
        colback=lightblue!30,   
        colframe=lightblue,     
        fonttitle=\bfseries,    
        coltitle=black,         
        sharp corners,          
        boxrule=0pt,            
        width=\columnwidth,     
        leftrule=1mm,           
        rightrule=0mm,          
        toprule=0mm,            
        bottomrule=0mm,         
    }
}

\lstdefinelanguage{json}{
  basicstyle=\ttfamily\scriptsize, 
  numbers=none,
  showstringspaces=false,
  breaklines=true, 
  breakatwhitespace=false, 
  frame=single,
  stringstyle=\color{red},
  morestring=[b]",
  moredelim=[s][\color{blue}]{\{}{\}},
  moredelim=[s][\color{green}]{[}{]},
}

\def\BibTeX{{\rm B\kern-.05em{\sc i\kern-.025em b}\kern-.08em
    T\kern-.1667em\lower.7ex\hbox{E}\kern-.125emX}}
\begin{document}



\title{Decoding Dependency Risks: A Quantitative Study of Vulnerabilities in the Maven Ecosystem}

\author{\IEEEauthorblockN{
Costain Nachuma ~~~~~~~~ Md Mosharaf Hossan ~~~~~~~~ Asif K. Turzo ~~~~~~~~ Minhaz F. Zibran}
\IEEEauthorblockA{\textit{Department of Computer Science, Idaho State University, Pocatello, ID, United States} \\
\{costainnachuma, mdmosharafhossan, asifkamalturzo, minhazzibran\}@isu.edu}
}



\maketitle

\begin{abstract}

This study investigates vulnerabilities within the Maven ecosystem by analyzing a comprehensive dataset of 14,459,139 releases. Our analysis reveals the most critical weaknesses that pose significant threats to developers and their projects as they look to streamline their development tasks through code reuse. We show risky weaknesses, those unique to Maven, and emphasize those becoming increasingly dangerous over time. Furthermore, we reveal how vulnerabilities subtly propagate, impacting 31.39\% of the 635,003 latest releases through direct dependencies and 62.89\% through transitive dependencies.
Our findings suggest that improper handling of input and mismanagement of resources pose the most risk. 
Additionally, Insufficient session-ID length in J2EE configuration and no throttling while allocating resources uniquely threaten the Maven ecosystem. We also find that weaknesses related to improper authentication and managing sensitive data without encryption have quickly gained prominence in recent years. These findings emphasize the need for proactive strategies to mitigate security risks in the Maven ecosystem.

\end{abstract}


\section{INTRODUCTION}
\vspace{-0.1cm}
The availability of modern software ecosystems like Maven Central plays a vital role in software development by providing developers with much-needed access to a vast repository of reusable libraries and tools. These ecosystems significantly simplify development by improving efficiency and enhancing software quality~\cite{b2, b5}. However, this convenience usually comes at a cost, that is, the introduction of substantial risks as vulnerabilities may be propagated through direct and transitive dependencies in the ecosystem and threaten the security and overall stability of dependent artifacts~\cite{b3}. The increased reliance on third-party libraries only intensifies these risks~\cite{b11}, especially in large ecosystems like Maven.


Despite the availability of some tools and techniques for vulnerability detection~\cite{b13,b14,b15}, the current approaches have been proven to fail often with regard to addressing the systemic nature of transitive vulnerabilities~\cite {b4, b6}. Recently, a critical vulnerability, CVE-2024-47, brings the risks that are inherent to dependency management to the spotlight within the Maven ecosystem~\cite{b7}. This signifies the urgent need for comprehensive studies that analyze vulnerabilities across dependency chains. 


The Mitre corporation~\cite{b9} tracks publicly disclosed cybersecurity vulnerabilities defined as Common Vulnerabilities and Exposures (CVE). The corporation also manages a list of Common Weakness Enumeration (CWE), which is defined as software or hardware weaknesses that can lead to exploitable security vulnerabilities. We aim to determine which weaknesses (CWE) are the most risky for the ecosystem. This finding would help developers and organizations address those CWEs proactively through focused coding standards or automated tooling. The Mitre corporation~\cite{b9} has released the top 25 most dangerous software weaknesses yearly since 2019. However, there is no such list for the Maven ecosystem. As a result, developers and organizations risk exploitation by considering only the weaknesses from the yearly top 25 lists from Mitre~\cite{b23}, which we refer to as the formal lists.

This study investigates risks posed by CWEs along with the propagation and impact of security vulnerabilities in the Maven ecosystem. By analyzing the trends in vulnerabilities and their distribution across dependency chains, we reveal hidden risks of modern ecosystems. Our findings aim to contribute to a deeper understanding of how vulnerabilities propagate and the challenges they pose to ecosystem security.

%
%
We explore the following research questions:

RQ1. How has the number of vulnerable releases changed over the years? 

RQ2. What types of weaknesses are most dangerous in the ecosystem? How do weaknesses in Maven differ from the formal list?

RQ3. Which weaknesses are gaining momentum?

RQ4. What proportion of releases are directly and transitively impacted by vulnerabilities?



To address RQ1, we look at the trend of vulnerable releases over the years. To answer RQ2, we analyze the risks involved with CWEs. We show 25 weaknesses that pose the most threat, present a top-level categorization, and identify four weaknesses unique to Maven ecosystem when compared to formal lists. To answer RQ3, we look at all the weaknesses in each of the last five years and show eight weaknesses that are becoming more dangerous. To address RQ4, we focus on the latest releases and examine their direct and transitive vulnerability status, revealing that 64.96\% are impacted by vulnerabilities. 
We provide a replication package~\cite{b1} comprising our scripts and the outputs used for our analysis.




\section{DATA COLLECTION}


To conduct our detailed analysis of the Maven ecosystem, we use the Goblin framework which offers a comprehensive dataset in the form of a Neo4j database. This dataset~\cite{b2} has vital information about artifacts, their releases, and associated dependencies. The version of the dataset we explore is dated August 30, 2024, enhanced with Weaver metrics including CVE mappings which we make use of to gain intricate insights into the structure and dynamics of the dependency network.

CVE data is linked to the Goblin dataset using the Weaver added-value nodes, matching CVE identifiers with affected versions. This ensures accurate mapping of vulnerabilities to releases and enables tracing their propagation through dependencies.

The dataset used in this study consists of 658,078 artifacts and 14,459,139 releases, of which 635,003 are identified as the latest releases. Additionally, 77,393 releases are associated with known vulnerabilities, providing insights into the security aspects of the Maven ecosystem.

\vspace{-0.1cm}

\section{ANALYSIS AND FINDINGS}
\vspace{-0.1cm}
\subsection{Addressing RQ1} 
\subsubsection{Approach} 
To address RQ1, we focus on the trend of the releases each year. Number of releases for each year since 2003 was calculated based on the release timestamp. Vulnerable releases are defined as those with CVE information. 
\subsubsection{Analysis} 
\begin{figure}[ht]
\centering
\includegraphics[width=0.8\linewidth]{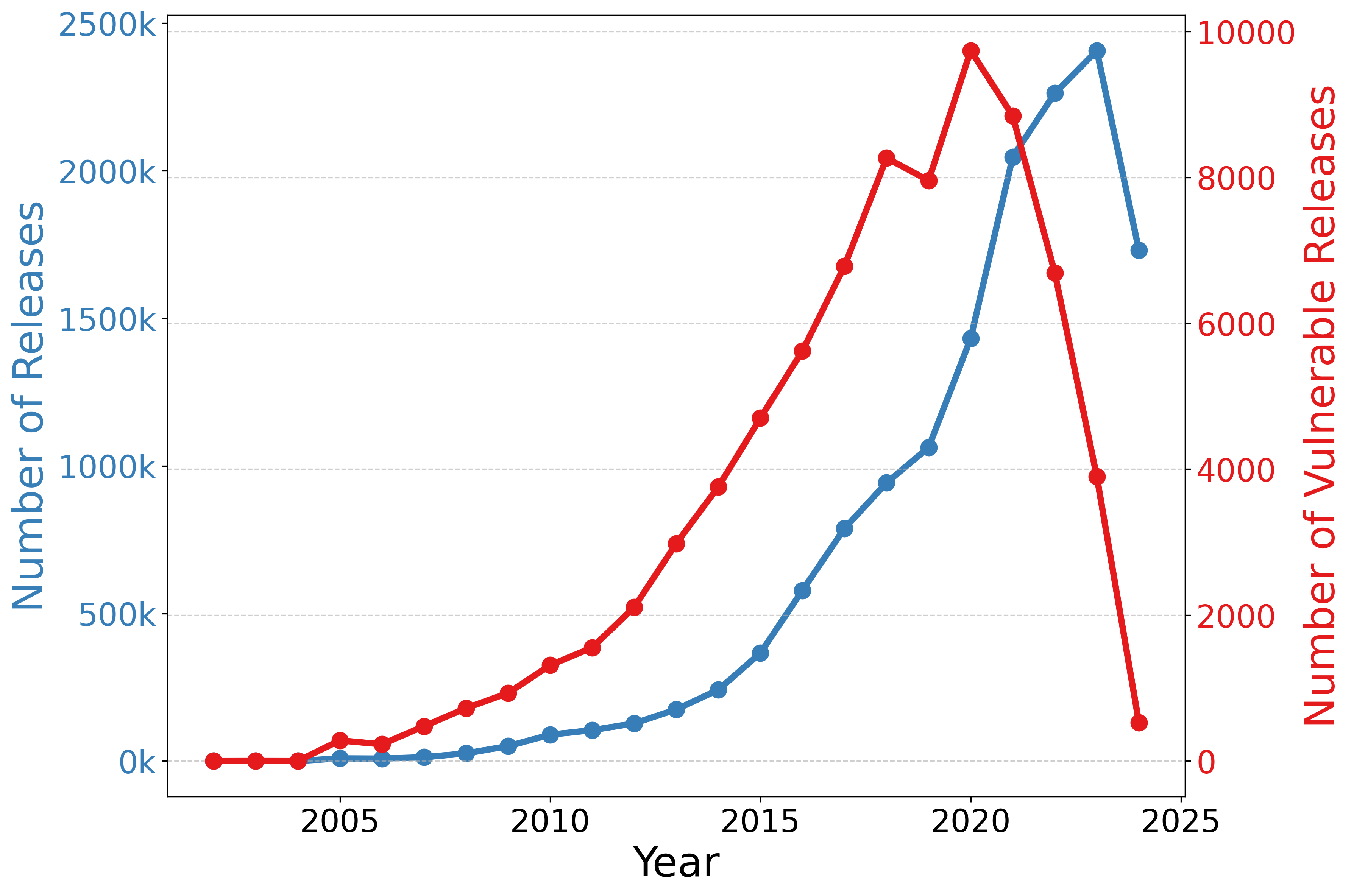}
\vspace{-0.2cm}
\caption{ Vulnerable Release Trend}
\vspace{-0.2cm}
\label{fig: Vulnerable Release Trend}
\end{figure}

In Figure~\ref{fig: Vulnerable Release Trend}, we show the trend of vulnerable releases.
With the growing number of releases, the number of vulnerable releases has also increased over the years. However, the percentage of vulnerable releases has slowly declined.
\vspace{-0.2cm}

Since the dataset only has partial data for 2024 (till August 30), numbers for this year represent a partial result.  In 2019, despite an increase in the number of releases, vulnerable releases decreased. Post-2020, the number of vulnerable releases showed a consistent decline.

\vspace{-.2cm}
\begin{tcolorbox}[boxrule=0.5pt, boxsep=-2pt, left=5pt, right=5pt]
\textbf{Answer to RQ1:} The percentage of vulnerable releases gradually declined over the years. After 2020, the number of vulnerable releases declined despite the increase in total number of releases.
\end{tcolorbox}



\subsection{Addressing RQ2}
\subsubsection{Approach}
We use the following metrics to find the most dangerous CWEs.


\textbf{Frequency ($\mathcal{F}$):} For frequency, we follow the same methodology~\cite{b10} adopted by the Mitre corporation~\cite{b9}. First, we count the number of times a CWE appears as a root cause of CVEs in the ecosystem for all CWEs. Let $\mathcal{F}(\omega)$ denote the Frequency of a CWE $\omega$. The counts are then normalized using the maximum and minimum counts in the list. Let $\mathcal{F}'(\omega)$ denote the normalized Frequency of a CWE $\omega$.



\textbf{Danger Value ($\mathcal{D}$):}  The danger value $\mathcal{D}$ of a CWE $\omega$, denoted as $\mathcal{D}(\omega)$, is calculated based on the severity of CVEs it led to. We adopted the methodology~\cite{b10} used by the Mitre corporation~\cite{b9}. The dataset lacks individual common vulnerability severity scores (CVSS) for each CVE. Therefore, average scores are assigned based on the severity ranges shown in Table \ref{table:cvss_score_range}. First, we count CVEs of each severity type for a $\omega$. Then, the counts are multiplied by the assigned Average CVSS scores ($\mathcal{S}$). The total of those represents $\mathcal{D} (\omega)$ for a $\omega$.






\begin{table}[h!]
\centering
\caption{CVSS v3.x Scores for Each Severity Type}
\vspace{-0.2cm}
\label{table:cvss_score_range}
\begin{tabular}{|l|c|c|}
\hline
\textbf{Severity} & \textbf{Range} & \textbf{Average Score ($\mathcal{S}$)} \\ \hline
None              & 0.0            & 0.0                             \\ \hline
Low               & 0.1-3.9        & 2                               \\ \hline
Moderate          & 4.0-6.9        & 5.45                            \\ \hline
High              & 7.0-8.9        & 7.95                            \\ \hline
Critical          & 9.0-10.0       & 9.5                             \\ \hline
\end{tabular}
\label{tab:cvss_scores}
\end{table}
\vspace{-0.4cm}
\begin{equation}
 \mathcal{D} (\omega) = \sum_{\text{Severity}} \text{count}(\omega) \times \mathcal{S}
 \vspace{-0.2cm}
\end{equation}

 Each $\mathcal{D}(\omega)$ is then normalized using the minimum and maximum values in the list. Let $\mathcal{D}'(\omega)$ denote the \emph{normalized} Danger Value of CWE $\omega$.
 
\textbf{Risk Score ($\mathcal{R}$)}: The level of risk incurred by a particular CWE $\omega$, denoted as $\mathcal{R}(\omega)$, is determined as follows: 
\vspace{-0.2cm}
\[
\mathcal{R}(\omega) = \mathcal{F}'(\omega) \times \mathcal{D}'(\omega) \times 100
\vspace{-0.2cm}
\]

We then show the top 25 most dangerous CWEs in the ecosystem based on the risk score. Next, we parse the category information from the Comprehensive Categorization for Software Assurance Trends~\cite{b8} and create a category to weakness map. Each category's total contribution equals the sum of the risk scores of its CWEs. We also report which CWEs consistently made Maven's top 25 list from 2019 to 2024 but were absent from formal lists~\cite{b23}.

\subsubsection{Analysis} 
Table \ref{table:cwe_danger_scores} shows that 4 CWEs tend to pose the most amount of threat in the Maven ecosystem. CWE-502: Deserialization of Untrusted data, CWE-79: Improper Neutralization of Input During Web Page Generation (`Cross-site Scripting'), CWE-400: Uncontrolled Resource Consumption, and CWE-20: Improper Input Validation. The Open Worldwide Security Project (OWASP)~\cite{b31} released guidelines to prevent and mitigate these weaknesses CWE-502:~\cite{b32}, CWE-79:~\cite{b33}, CWE-400:~\cite{b30}, CWE-20:~\cite{b35}~\cite{b36}.

\begin{table}[H]\centering
\caption{Risk Scores of Top CWEs}
\vspace{-.2cm}
\label{table:cwe_danger_scores}
\renewcommand{\arraystretch}{1.1}
\begin{tabular}
{|l@{ }>{\raggedleft\arraybackslash}p{1cm}@{ }|@{}c
|l@{ }>{\raggedleft\arraybackslash}p{1cm}@{ }|@{}c
|l@{ }>{\raggedleft\arraybackslash}p{1cm}@{ }|}
\cline{1-2} \cline{4-5} \cline{7-8}
\textbf{CWE} & \textbf{$\mathcal{R}$} & & \textbf{CWE} & \textbf{$\mathcal{R}$} & & \textbf{CWE} & \textbf{$\mathcal{R}$} \\
\cline{1-2} \cline{4-5} \cline{7-8}
CWE-502 & 191.40 & & CWE-287 & 11.49 & & CWE-668 & 4.03   \\
CWE-79  & 140.16 & & CWE-444 & 9.45  & & CWE-284 & 3.58  \\
CWE-20  & 133.05 & & CWE-863 & 9.34  & & CWE-295 & 3.57 \\
CWE-400 & 102.24 & & CWE-918 & 6.17  & & CWE-835 & 3.51  \\
CWE-22  & 61.56  & & CWE-269 & 4.81  & & CWE-601 & 3.46 \\
CWE-200 & 44.33  & & CWE-83  & 4.65  & & CWE-917 & 3.00  \\
CWE-611 & 20.29  & & CWE-352 & 4.41  & & CWE-77  & 2.86  \\
CWE-94  & 16.63  & & CWE-74  & 4.29  & & CWE-89  & 2.65 \\
CWE-770 & 15.53  & &         &  & &        &       \\

\cline{1-2} \cline{4-5} \cline{7-8}
\end{tabular}
\vspace{-0.1cm}
\end{table}

\begin{figure}[ht]
 \vspace{-0.2cm}
\centering
\includegraphics[scale=0.25]{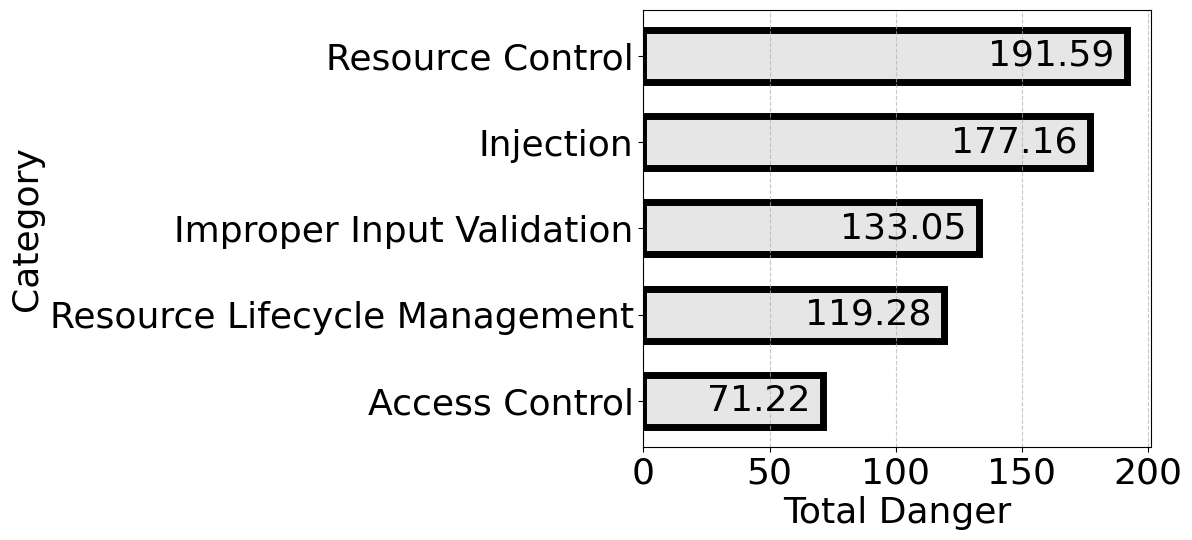}
\vspace{-0.2cm}
\caption{Top 5 CWE Mappings by Category}
\vspace{-0.2cm}
\label{fig: Top 5 CWE Mappings by Category}
\end{figure}

Based on the categorization~\cite{b8}, Figure \ref{fig: Top 5 CWE Mappings by Category} shows five categories that stand out. two involve input mishandling, two resource mismanagement, and one improper access control.

Maven's unique weakness calculation indicates that it differs from formal lists by 4 CWEs. Those are CWE-6: J2EE Misconfiguration: Insufficient Session-ID Length, CWE-40: Path Traversal: \texttt{\textbackslash\textbackslash UNC\textbackslash share\textbackslash name\textbackslash} (Windows UNC Share), CWE-770: Allocation of Resources Without Limits or Throttling, CWE-86: Improper Neutralization of Invalid Characters in Identifiers in Web Pages. Among them, CWE-6 and CWE-770 pose the highest threat, with 34.63 and 21.77 average risk scores. OWASP~\cite{b31} released guidelines to properly manage session to avoid CWE-6~\cite{b27}. CWE-770 is susceptible to different flooding and denial of service (DoS) attacks. OWASP~\cite{b31} also discussed ways to avoid CWE-770 and DoS attacks~\cite{b28,b29,b30}.




\vspace{-.2cm}
\begin{tcolorbox}[boxrule=0.5pt, boxsep=-2pt, left=5pt, right=5pt]


\textbf{Answer to RQ2:} Improper input handling and resource mismanagement poses the most risks, with CWE-6 and CWE-770 uniquely threatening Maven Central.
\end{tcolorbox}
\vspace{-.3cm}

\subsection{Addressing RQ3}
\subsubsection{Approach} 

To respond to RQ3, we calculate the risk score for all CWEs each year from 2020 to 2024 based on the same methodology we followed in RQ2. We then proceeded to identify which CWEs either maintained the same rank or showed a consistent upward trend in rank.
\subsubsection{Analysis} 

\begin{figure}[ht]
\vspace{-0.4cm}
\centering
\includegraphics[width=1\linewidth]{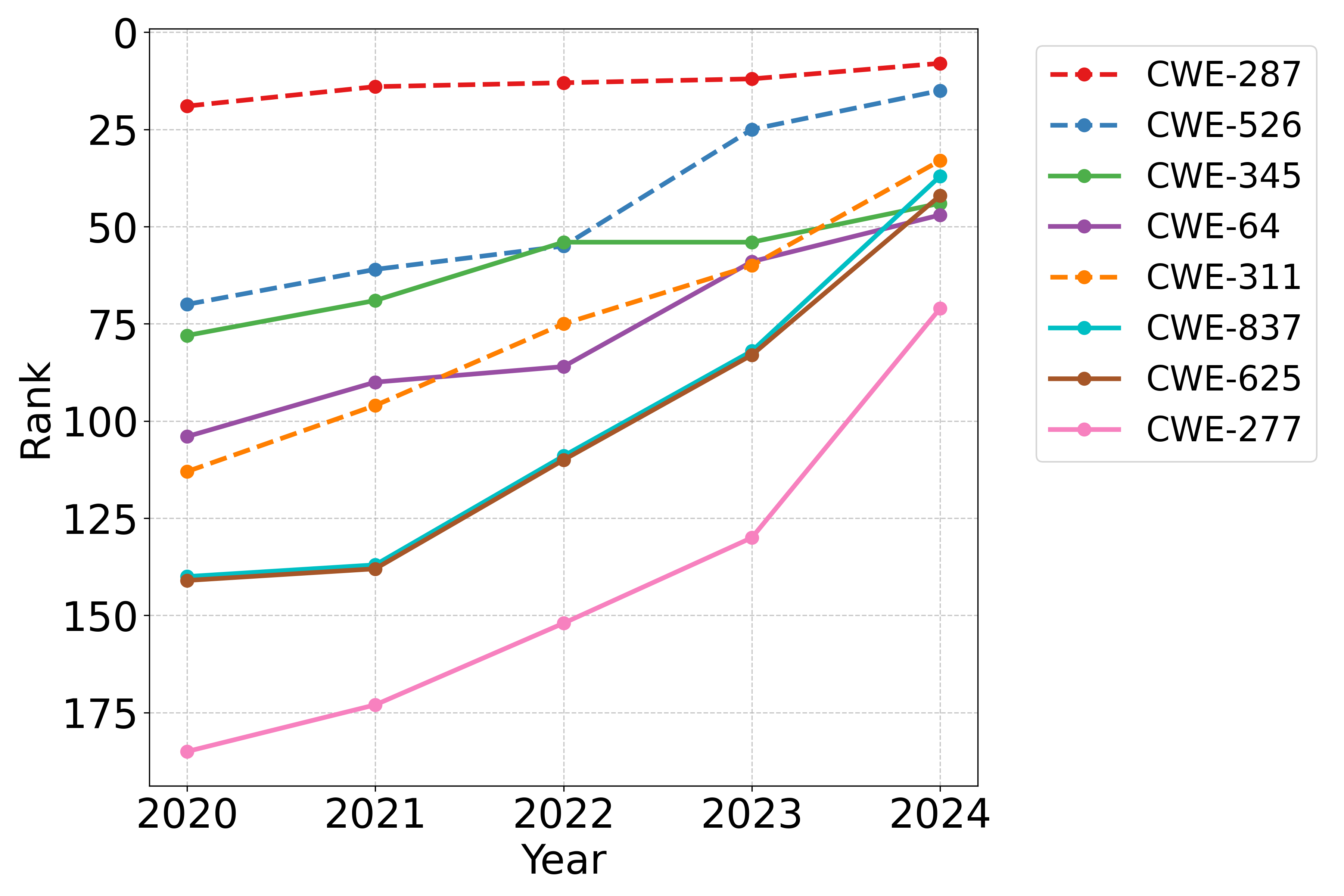}
\vspace{-0.6cm}
\caption{Upward Trend CWEs}
\label{fig: Upward Trend CWEs}
\end{figure}

Figure \ref{fig: Upward Trend CWEs} shows the trend of eight CWEs that have consistently maintained or improved their rank over the last five years. Among the top 25 dangerous CWEs, CWE-287: Improper Authentication has consistently risen through the ranks over the years. CWEs are structured in a tree-like shape with the following order: Pillar, Class, Base, and Variant. CWE-287 is a class-level weakness. Its offsprings CWE-288, CWE-645, CWE-613, CWE-640, and CWE-526 have also had an upward trend.

One interesting trend here is the rise of CWE-526: Cleartext Storage of Sensitive Information in an Environment Variable and CWE-311: Missing Encryption of Sensitive Data. In CWE structure, CWE-526 is a offspring of CWE-311. In 2024, they rank 15 and 33 having started from 70 and 113 respectively. This underscores the urgent need for stronger safeguards through encryption against sensitive data exposure. Ideas for handling sensitive data in Maven is discussed in a apache wiki article~\cite{b24}. Ways like using GPG agents, having stronger filesystem permissions or using open source libraries~\cite{b25,b26} is discussed. OWASP~\cite{b31} also has guidelines for preventing CWE-526 and CWE-311~\cite{b37,b38}.

\vspace{-.2cm}
\begin{tcolorbox}[boxrule=0.5pt, boxsep=-2pt, left=5pt, right=5pt]
\textbf{Answer to RQ3:} Weaknesses related to managing sensitive data without encryption have become alarmingly risky over the last 5 years.
\end{tcolorbox}

\subsection{Addressing RQ4}
\subsubsection{Approach} 
To answer RQ4, we focus on the latest releases in the Maven ecosystem. This approach ensures computational feasibility and guarantees that all artifacts are represented in the evaluation. Analyzing transitive dependencies for over 14 million releases is computationally costly and potentially introduces noise from outdated releases that are no longer significant for active projects.

\paragraph{Extracting Vulnerable Releases}  
We begin by extracting all releases with a known vulnerability from the ecosystem. These vulnerabilities are identified through their relationship to the added values representing CVEs in the dataset.

\paragraph{Extracting Latest Releases and Dependencies}  
To isolate the latest releases, we identify the most recent release timestamp for each artifact in the dependency graph and analyze their direct and transitive dependencies. For transitive dependencies, we limit the analysis to a maximum depth of three levels. This constraint balances computational cost and practical relevance, as prior work~\cite{b3} suggests that significant relationships predominantly exist within 2–3 levels. This process is computationally expensive, requiring approximately 21 hours on a system with 64 GB RAM and 24 CPU cores.

\paragraph{Mapping Vulnerabilities to Releases}  
We map vulnerabilities to each latest release using CVE values extracted from the CVE added-values nodes in the dataset. This process evaluates vulnerabilities in both direct and transitive dependencies across cascading levels.

\paragraph{Metrics for the Ecosystem Analysis}  
We use the following metrics for the impact analysis:


Proportion of Directly Impacted Releases (\(P_d\)):
The proportion of directly impacted releases is calculated as the percentage of the latest releases (\(L\)) affected by at lease one CVE in their direct dependencies (\(D\)): $P_d = \frac{D}{L} \times 100$.


Proportion of Transitively Impacted Releases (\(P_t\)): 
The proportion of transitively impacted releases (\(P_t\)) is calculated as the percentage of latest releases (\(L\)) affected by vulnerabilities in transitive dependencies (\(T\)) across Levels 1, 2, and 3: $P_t = \frac{T}{L} \times 100$.



\subsubsection{Analysis} 
\begin{figure}[ht]
\centering
\vspace{-0.3cm}
\includegraphics[scale=0.25]{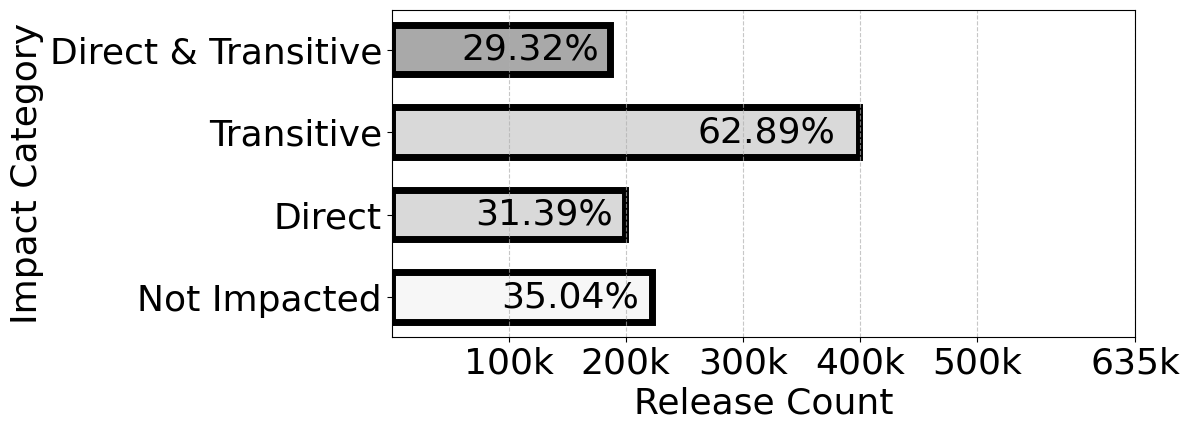}
\caption{Vulnerability Impact Analysis Results}
\label{fig:impact}
\vspace{-0.3cm}
\end{figure}

 Fig.~\ref{fig:impact} reveals a striking contrast in the way direct and transitive vulnerabilities impact the releases within the Maven ecosystem.

\vspace{5pt}
\paragraph{Direct Vulnerabilities: A Manageable Frontier?}

About 31.39\% of the latest releases are impacted by direct vulnerabilities. While significant, this proportion represents a manageable challenge as direct vulnerabilities are linked to immediate dependencies, hence making them easier to detect and mitigate. This emphasizes the importance of strong security practices at the release level, including vetting dependencies and leveraging automated detection tools. Addressing direct vulnerabilities at the source has great promise in reducing the attack surface.

\vspace{-0.1cm}
\paragraph{Transitive Vulnerabilities: The Hidden Threat}

An alarming 62.89\% of the latest releases are impacted by vulnerabilities propagated through three levels of transitive dependencies. Limiting the analysis to two levels of dependency chains reduces this to 60.20\%, a negligible difference of 2.69\%. As opposed to direct vulnerabilities, transitive ones are harder to trace and mitigate because they are spread across multiple layers. They often involve rarely updated or unmaintained dependencies deep in the graph. These deeply hidden risks amplify the impact of a single vulnerability in several artifacts. This underscores the interconnected fragility of the ecosystem and the need for tools to reveal hidden risks.

\vspace{-0.1cm}
\paragraph{Implications for Ecosystem Security}
Out of 635,003 total latest releases, only 35.04\% of the latest releases are free from direct and transitive vulnerabilities. In contrast, 29.32\% of the releases are affected by both types of vulnerabilities compounding the challenge. This distinction calls for a shift in how developers and maintainers manage dependencies, as traditional practices focusing solely on direct dependencies are insufficient. To minimize the uncovered vulnerabilities, we propose ecosystem tools consider integrating deeper insights into transitive dependencies offering proactive alerts and recommendations to developers. In addition, collaborative efforts among library maintainers should be fostered to safely deprecate unmaintained dependencies and enforce stricter security checks for widely used artifacts. Furthermore, incorporating automated updates for vulnerable dependencies, especially for transitive ones would significantly improve the ecosystem security.
\vspace{-.2cm}
\begin{tcolorbox}[boxrule=0.5pt, boxsep=-2pt, left=5pt, right=5pt]
\textbf{Answer to RQ4:} 
31.39\% of the latest releases are directly impacted by vulnerabilities while 62.89\% are affected by transitive vulnerabilities highlighting the critical need for tools and collaboration to mitigate systemic risks.
\end{tcolorbox}
\vspace{-0.3cm}



\section{THREATS TO VALIDITY}
\vspace{-.1cm}
Our top 25 most dangerous weakness list relies on the risk score found for each CWE. To find that we used the average CVSS score for a severity and used that value for all CVEs of that severity. Using individual CVSS scores would have provided a more nuanced rank. 

Limiting RQ4 analysis to the latest releases for computational feasibility and relevance, we may have omitted valuable insights from older active releases.

Also, the findings of this study might only be applicable to the Maven ecosystem and might not be applicable to other ecosystems. 





\vspace{-0.1cm}
\section{RELATED WORK}
\vspace{-0.1cm}
Decan et al.~\cite{b3} studied security vulnerabilities in the npm ecosystem showing how the number of reported vulnerabilities has increased over time showing the importance of proactive vulnerability management. Similarly, the National Vulnerability Database~\cite{b4} has highlighted significant trends in vulnerabilities affecting modern software systems, including Maven. Prior studies also proposed metrics to identify security vulnerability~\cite{b39,b40}, identified which vulnerability remains hidden~\cite{b41,b42}, and which vulnerabilities are identified in the code review process~\cite{b43,b44,b45}.

Ruohonen et al.~\cite{b21} ranked a list of weaknesses based on frequency while empirically analyzing vulnerabilities in the Python packages for web applications. Another study conducted by Alfadel et al. \cite{b22} showed the yearly trend of release and vulnerable release as part of empirically finding security vulnerabilities in the Python packages. Luo et al.~\cite{b20} investigated the threats posed by upstream vulnerabilities to downstream projects in the Maven ecosystem. Their results revealed the complex dynamics of dependency relationships and stressed the need for urgent vulnerability mitigation strategies to protect downstream artifacts. Recently Rabbi et al.~\cite{b19} examined SBOM generation tools in the npm ecosystem highlighting gaps in dependency representation and vulnerability identification. Their findings align with our focus on analyzing direct and transitive vulnerabilities in the maven ecosystem.

\vspace{-0.2cm}


\vspace{-0.1cm}
\section{CONCLUSION }
\vspace{-0.1cm}
This study explores the vulnerabilities within the Maven ecosystem focusing on risks related to dependencies that threaten software security and stability. By analyzing over 14 million releases we identify critical weaknesses that include those unique to the Maven ecosystem. The findings also highlight the disparate impact of transitive vulnerabilities which is nearly double the number of direct vulnerabilities. Our research emphasizes the importance of proactive dependency management and the development of tools that offer robust handling of vulnerabilities within the Maven ecosystem.


In future, we plan to identify which CWEs have tendency to propagate further in the dependency chain and incorporate this factor in evaluating the risk score of a CWE.

\section*{Acknowledgement}
This work is supported in part by the ISU-CAES (Center for Advanced Energy Studies) 
Seed Grant at the Idaho State University, USA.

\end{document}